\documentclass[reprint, superscriptaddress, aps, jcp]{revtex4-1}

\usepackage[T1]{fontenc}
\usepackage{graphicx}
\usepackage{xfrac}

\usepackage[
  unicode,
  pdfusetitle,    % title, authors and date as pdf attributes
  pdfcreator={},
  pdfproducer={},
]{hyperref}

\usepackage{mathtools}

\let\theta\vartheta
\newcommand*{\dd}{\mathrm{d}}
\newcommand*{\pp}{\mathrm{p}}
\newcommand*{\bu}{\textbf{u}}

\newcommand*{\br}{\textbf{r}}

\usepackage{xcolor}
\colorlet{myred}{red!80!black}

\begin{document}

%%%%%%%%% TITLE SECTION %%%%%%%%%%
\title{Unusual geometric percolation of hard nanorods in the uniaxial nematic liquid crystalline phase}
\author{Shari P. Finner}
\email{s.p.finner@tue.nl}
\affiliation{Department of Applied Physics, Eindhoven University of Technology, P.O. Box 513,
3500 MB Eindhoven, The Netherlands}
\author{Arshia Atashpendar}
\email{arshia.atashpendar@physik.uni-freiburg.de}
\affiliation{Physikalisches Institut, Albert-Ludwigs-Universit\"at Freiburg, 79104 Freiburg, Germany}
\author{Tanja Schilling}
\email{tanja.schilling@physik.uni-freiburg.de}
\affiliation{Physikalisches Institut, Albert-Ludwigs-Universit\"at Freiburg, 79104 Freiburg, Germany}
\author{Paul van der Schoot}
\email{p.p.a.m.v.d.schoot@tue.nl}
\affiliation{Department of Applied Physics, Eindhoven University of Technology, P.O. Box 513,
3500 MB Eindhoven, The Netherlands}
\affiliation{Institute for Theoretical Physics, Utrecht University, Princetonplein 5, 3584 CC Utrecht, The Netherlands}

\begin{abstract}
	We investigate by means of continuum percolation theory and Monte Carlo simulations how spontaneous uniaxial symmetry breaking affects geometric percolation in dispersions of hard rod-like particles.
	If the particle aspect ratio exceeds about twenty, percolation in the nematic phase can be lost upon adding particles to the dispersion.
	This contrasts with percolation in the isotropic phase, where a \textit{minimum} particle loading is always required to obtain system-spanning clusters.
	For sufficiently short rods, percolation in the uniaxial nematic mimics that of the isotropic phase, where the addition of particles always aids percolation.
	For aspect ratios between twenty and infinity, but not including infinity, we find re-entrance behavior: percolation in the low-density nematic may be lost upon increasing the amount of nanofillers but can be re-gained by the addition of even more particles to the suspension.
	Our simulation results for aspect ratios of 5, 10, 20, 50 and 100 strongly support our theoretical predictions, with almost quantitative agreement.
	We show that a new closure of the connectedness Ornstein-Zernike equation, inspired by Scaled Particle Theory, is more accurate than the Lee-Parsons closure that effectively describes the impact of many-body direct contacts.
\end{abstract}

\maketitle

%%%%%%%%%%%%%%%%%%%%%%%%%%%

	\section{Introduction}
		The mechanical, electrical or thermal properties of common engineering plastics can be enhanced by adding sufficient amounts of nanoparticles during the fluid stages of the material processing.\,\cite{KoningCNTbook, Torquatobook, Foygel2005, Ackermann2016, Grossiord2008, Mutiso2015, Thomassin2013}
		Such an enhancement of material properties is desirable for many nanotechnological applications in optoelectronics, electromagnetic interference shielding and photovoltaics.\,\cite{DresselhausCNTbook,Mutiso2015,Grossiord2006, Grossiord2008, Deng2009,Thomassin2013} 
		Often, there is a critical filler loading above which these properties are improved strongly nonlinearly with the number of particles added.
		This critical concentration is called the percolation threshold and is caused by the particles forming a system-spanning network in the material.
		A common engineering goal is to keep the percolation threshold as low as possible in order to preserve other qualities of the polymeric host material of choice, such as optical transparency and ease of processing.
		For instance, a polymer-based thin-film transparent electrode requires a minimal loading of conductive fillers, as the fillers deteriorate the optical transparency of the films.
				
		It stands to reason that, for an efficient and rational design of novel materials, it is crucial to understand the key factors influencing the formation of particle clusters in nanocomposites.
		For slender rod-like filler particles, it has been shown theoretically and by means of computer simulations that the percolation threshold should scale inversely proportionally to their aspect ratio, and experiments seem to support this.\,\cite{Nigro2013,Kyrylyuk2008,Deng2009,Ambrosetti2010,Otten2011,Mutiso2015,Schilling2015,Drwenski2017}
		As a result, strongly elongated fillers such as carbon nanotubes percolate at very low volume fractions, making them particularly suited for applications where the transparency of the host material needs to be preserved. \cite{Mutiso2015, Foygel2005, Grossiord2008, Ackermann2016}
		The underlying assumption in many of these studies is that the filler particles are oriented isotropically, which in practice need not be the case due to, e.g., processing or confinement in a thin film. \cite{KoningCNTbook, Moradi2015, Finner2018}
		In fact, we expect the percolation threshold to be close to the critical concentration at which long rod-like particles spontaneously form a nematic liquid crystal in an isotropic host fluid. \cite{Otten2012, Finner2019PRL}
		At high enough packing fraction, the particles in the isotropic phase run out of space due to excluded volume interactions and align along a common axis, \textit{i.e.}, the uniaxial nematic phase becomes the stable phase. \cite{Onsager1949, Vroege1992}
		This also implies that particles necessarily become relatively close to their immediate neighbors in the dispersion, suggesting the vicinity of the percolation transition.

		In fact, the percolation transition may even be preempted by the liquid crystal phase transition.
		At this moment, very little is known about percolation of elongated particles in liquid crystalline phases.
		What is known is that, in the limit of infinite aspect ratio of cylindrical particles that interact via a hard-core steric repulsion, the formation of a system-spanning cluster in the nematic phase is nearly independent of the filler fraction. \cite{Finner2019PRL}
		Whether or not the average cluster is system-spanning in the nematic depends, by and large, on the physics defining particle connections rather than the density.
		For electrical conductivity, for instance, particle connectivity is mainly determined by the average tunneling distance of charge carriers between adjacent nanoparticles through the polymer matrix. \cite{Kyrylyuk2008, Sherman1983, Balberg2009, KoningCNTbook, Ambrosetti2010, Mutiso2015, Lagerwall2008review, Lagerwall2016book}
		This is why electrical percolation is often viewed as geometric percolation, where the so-called ``connectivity range'' is a criterion for the maximum surface-to-surface distance between two particles in order to still be considered connected.\cite{Kyrylyuk2008, Balberg2009, Otten2011}
		In this paper, we show by means of connectedness percolation theory and Monte Carlo simulations that the critical connectivity range in the nematic is a weak but complex function of the concentration, and becomes only modestly sensitive to the aspect ratio in the high-density nematic phase and near the melting transition to the isotropic phase.
		This is quite unusual, given that the isotropic-to-nematic transition depends very strongly on the particle aspect ratio. \cite{Nigro2013,Kyrylyuk2008,Deng2009,Ambrosetti2010,Otten2011,Mutiso2015,Schilling2015,Drwenski2017}
		Consequently, tuning the percolation threshold in the uniaxial nematic is for all intents and purposes restricted to modifying the tunneling range, which in practice is set by the dielectric constant of the medium and the type and properties of the filler particles.  \cite{Kyrylyuk2008, Sherman1983, Balberg2009, KoningCNTbook, Ambrosetti2010, Mutiso2015, Lagerwall2008review, Lagerwall2016book}
		
		It is important to stress that we find the impact of spontaneous alignment to be of a fundamentally different nature than that of forced alignment by an external orienting field.
		Imposed orienting fields such as electric, magnetic and flow fields tend to increase the percolation threshold, because alignment leads to larger surface-to-surface distances. \cite{Balberg1984, Balberg1984_2, Kale2016, Otten2012, Finner2018, Chatterjee2014, White2009, Deng2009, Kumar2016}
		In the nematic, increased alignment is induced by an increase in density, which almost exactly compensates for the alignment effect, at least so in the limit of infinite particle aspect ratio.\cite{Finner2019PRL}
		As we show in this paper, the compensation effect is only partial for rods of finite aspect ratio.
		This is due to the contributions of configurations where the spherical end-caps connect with each other or with the cylindrical body of the particles.
		These configurations become increasingly important the shorter the rods are, in which case also higher-order body interactions come into play.
		We find that if the rods are sufficiently short, percolation in the uniaxial nematic resembles that in the isotropic phase: the addition of particles always leads to an increase in cluster size.
		For infinitely large aspect ratios this is no longer the case, and percolation can be lost with increasing concentration. \cite{Finner2019PRL}
		For aspect ratios between about twenty and infinity, we find re-entrance behavior, where percolation in the low-density nematic may be lost upon increasing the amount of nanofillers but can be re-gained by the addition of even more particles to the suspension.
		
		The remainder of this paper is organized as follows.
		In Section \ref{sec:theory}, we first outline our theoretical framework, which hinges on a generalized Onsager theory for the orientational distribution function of hard spherocylinders, and on connectedness percolation theory describing the clustering of particles.
		To address the percolation problem in the nematic analytically, we propose in Section \ref{sec:vartheory} a variational percolation theory, the outline of which we presented in a recent publication. \cite{Finner2019PRL}
		In Section \ref{sec:gammas}, we discuss two closures for the percolation equation, which take into account finite particle aspect ratios by effectively renormalizing the two-particle excluded volume and contact volume.
		The latter describes the volume that the center of mass of a particle must physically be in to make contact with a test particle. \cite{Balberg1984, Balberg1984_2}
		Details of our numerical and simulation methods are given in Section \ref{sec:numericsMC}.
		In Section \ref{sec:results}, we present percolation thresholds in terms of the critical connectivity range as a function of the filler fraction for a range of aspect ratios, and compare our theoretical predictions with the results of our Monte Carlo simulations.
		Section \ref{sec:dis} summarizes our main findings in the context of previously published results and provides suggestions for future directions.

%%%%%%%%%%%%%%%%%%%%%%%%%%%

	\section{Theoretical Model \label{sec:theory}}
		We model our nanoparticles as straight spherocylinders, that is, cylindrical bodies of diameter $D$ and length $L$, donned with hemispherical end caps of the same diameter.
		They interact via a harshly repulsive potential that is infinite if two particles overlap and zero if they do not.
		The hard, impenetrable core of a rod is centered inside of a penetrable, spherocylindrical contact shell of the same length, with diameter $D+\lambda$, where $\lambda$ denotes the connectivity range.
		Two rods are considered to be \textit{directly} connected if their contact shells overlap, \textit{i.e.}, if their surface-to-surface distance is smaller than $\lambda$.
		For electrical percolation in polymeric nanocomposites, $\lambda$ is equivalent to an effective tunneling distance of charge carriers through the host medium. \cite{Otten2011, Balberg2009, Shklovskii2006, Ambrosetti2010,Atashpendar2018}
		If the solution is aqueous and charge-stabilized, then the charge carriers are mobile ions, and $\lambda$ must arguably correspond to the Debye length.\cite{Vigolo2005, Finner2019PRL}
		It is important to note that the connectivity range, $\lambda$, is a material property determined by both the nanofiller and the host fluid, and is in principle controllable.
		For instance, one would expect the tunneling distance to depend on the Fermi energy of the filler and the dielectric constant of the medium. \cite{Kyrylyuk2008}
		It stands to reason that the interface between the two should also impact on the value of $\lambda$. \cite{Kashfipour2018}
		From now on, we shall treat $\lambda$ as an adjustable parameter.
		
		Within our model, the orientation of a particle's main body axis vector, $\bu$, is defined relative to the nematic director, which we choose to be the $z$-axis in our Cartesian coordinate system, so that $\bu^T= (\sin \theta \cos \phi, \sin \theta \sin \phi, \cos \theta )$, with $\theta$ and $\phi$ the usual polar and azimuthal angles.
		In the isotropic phase, where there is no director, the $z$-axis is an arbitrary axis.
		The orientational distribution function we write as $\psi (\bu)$.
		In the isotropic phase, $\psi (\bu) = 1/4\pi$, while in the nematic phase, it depends on the number density, $n$, of the nanofillers due to their interactions. 
		The orientational distribution in the nematic phase obeys cylindrical and inversion symmetry, so that $\psi (\bu) = \psi (- \bu)$ and $\psi (\bu)=\psi (\theta) = \psi (\pi - \theta)$.
		Clusters of particles in the isotropic and the nematic phase are described by a two-body distribution function called the pair connectedness function $P(\br, \bu, \bu')$, which is a function of the relative position, $\br$, of two particles and of their orientations $\bu$ and $\bu'$.
		The pair connectedness function also depends on the number density, $n$, and on the orientational distribution function, $\psi (\bu)$.
		Hence, in order to study cluster formation in the nematic phase, we first need to calculate the orientational distribution function.
		For this, we make use of a renormalized Onsager theory. \cite{Onsager1949, Parsons1979, Lee1987, TuinierBook, Cotter1977}
		
		The original Onsager theory is based on the second virial approximation of the free energy, which is written as a functional of the orientational distribution function.
		It is believed to become exact in the limit of infinite aspect ratio of the particles that we define as $L/D$.\cite{Onsager1949,Odijk1986}
		(Note that in some works the aspect ratio is defined as $L/D + 1$ to account for the hemispherical end caps.)
		The theory becomes quantitative for aspect ratios in excess of a few hundred. \cite{Frenkel1987, Lee1987} % at L/D=100: B3/B2^2=9%. At 1000: 1.5%
		To deal with the impact of a finite aspect ratio, we need to explicitly account for higher order virials, or alternatively make use of a suitable renormalization that incorporates them approximately.
		For simplicity, we choose to do the latter.
		Formally, we can write the Onsager equation for the orientational distribution function in its generalized form as		
		\begin{align}
			\ln \psi (\bu) &= k  + 2 n \langle \hat{C}(\bu, \bu') \rangle', \label{GeneralOnsager}
		\end{align}
		where $k$ is a Lagrange multiplier enforcing normalization of the orientational distribution function and $\langle \cdots \rangle' = \int \dd \bu' (\cdots) \psi(\bu')$ an angular average. \cite{Onsager1949, Vroege1992, Lee1987, TuinierBook}
		$\hat{C}(\bu, \bu') = \int \dd \br C(\br, \bu, \bu')$ is the volume integral of the direct correlation function $C(\br, \bu, \bu')$.
		Here we tacitly take as a reference state a gas with $n \rightarrow 0$. \cite{Hansen}
		
		Within the second virial approximation, $C(\br, \bu, \bu')= f(\br, \bu, \bu')$, with $f(\br, \bu, \bu') = \exp[ - \beta U(\br, \bu, \bu')] -1 $ the Mayer function. \cite{Hansen}
		Here, $\beta=1/k_\mathrm{B}T$ denotes the reciprocal thermal energy, and $U(\br, \bu, \bu')$ the interaction potential between two rods at relative position $\br$ and with the orientations $\bu$ and $\bu'$.
		By interpolating the Onsager equation of state of infinitely slender hard rods and the Carnahan-Starling equation of state for hard spheres, Lee and Parsons rescaled the excluded volume between two particles, which in our prescription may be expressed as $C(\br, \bu, \bu') = \Gamma_\text{LP}(n, L, D) f(\br, \bu, \bu')$. \cite{Parsons1979, Lee1987}
		Here, $\Gamma_\text{LP}(n, L, D)$ is a renormalization factor, which approximately accounts for higher than two-body contacts.
		An alternative treatment of higher-order body interactions is given by the framework of Scaled Particle Theory, \cite{TuinierBook} in which case we have $C(\br, \bu, \bu') = \Gamma_\text{SPT}(n, L, D) f(\br, \bu, \bu') + \Omega_\text{SPT}(\br, n, L, D)$ with $\Omega_\text{SPT}(\br, n, L, D)$ an additional renormalization term, see Appendix \ref{app:A}.
		This additional term does not depend on the particle orientation, and therefore does not impact upon the orientational distribution function, as it only rescales the Lagrange multiplier $k$ in Eq. \eqref{Onsager}.
		This means that the renormalized Onsager equation within Scaled Particle Theory (SPT) attains the same form as that within the Lee-Parsons (LP) prescription.
		For both approaches, $\Gamma(n, L, D)\to \Gamma(\phi, L/D) \geq 1$, with $\phi = n \pi D^2 [3L + 2D]/12$ the volume fraction of particles. \cite{Lee1987, TuinierBook}
		We provide explicit expressions for the scaling factors $\Gamma_\text{LP}(\phi, L/D)$ and $\Gamma_\text{SPT}(\phi, L/D)$ in Section \ref{sec:gammas} of this article.
		
		The volume integral of the Mayer function, $\hat{f}(\bu,\bu')$, the absolute value of which is equal to the excluded volume of two particles, obeys \cite{Onsager1949}
		\begin{align}
			- \hat{f} (\bu,\bu') = 2 L^2 D |\bu \times \bu' |+ 2 \pi L D^2 + \frac{4\pi}{3} D^3. \label{vex}
		\end{align}		
		With this, the renormalized Onsager equation takes the form
		\begin{align}
			\ln \psi (\bu) &= k -  \frac{8}{\pi} \Gamma(\phi, L/D) c  \int \dd \bu' \psi(\bu') |\bu \times \bu'|, \label{Onsager}
		\end{align}
		noting that all angle-independent terms can be absorbed in the Lagrange multiplier that we again denote by $k$.
		Relevant to the problem, we introduced the dimensionless concentration scale $c = n \pi L^2 D / 4$.
		For any given scaling factor $\Gamma(\phi, L/D)$, we can obtain the self-consistent orientational distribution function $\psi(\bu)$ by solving Eq.\,\eqref{Onsager} numerically by means of recursive iteration. \cite{vanRoij2005}
		A useful \textit{analytical} approximant for the orientational distribution, which we shall be using in our analytical percolation theory, is the Gaussian distribution put forward by Odijk.\,\cite{Odijk1986review, Vroege1992}
		To leading order in $c \gg 1$, it reads
		$\psi(\bu)=\psi(\theta)=\alpha \exp (- \alpha \theta^2/2) /4 \pi $ for $0 \leq \theta \leq \pi/2$
and $\psi(\pi - \theta)$ for $\pi/2 \leq \theta \leq \pi$, with $\alpha = 4c^2 \Gamma^2(\phi) / \pi$.\,\cite{Odijk1986review, Vroege1992}
		The Gaussian approximation is most accurate deep in the stable nematic phase.
		Nonetheless, it also gives a reasonable estimate of the orientational distribution close to the melting transition to the isotropic phase. \cite{Vroege1992, Odijk1986review}
		
		\section{Continuum Percolation Theory\label{sec:percTheory}}
		In order to study the clustering of particles, it is useful to separate all two-body distribution functions into contributions from connected and disconnected particles. 
		For instance, the radial distribution function, $g(\br, \bu, \bu') = P(\br, \bu, \bu') + D(\br, \bu, \bu')$, can be written as the sum of the pair connectedness function, $P(\br, \bu, \bu')$, and the pair blocking function, $D(\br, \bu, \bu')$. \cite{Coniglio1977}
		The former describes the unnormalized probability that two particles with orientations $\bu$ and $\bu'$ at relative position $\br$ are part of the same cluster, while the latter is the contribution of \textit{disconnected} particles to the total correlation function, \textit{i.e.}, particles that are neither directly nor indirectly connected.
		The pair connectedness function then obeys the connectedness Ornstein-Zernike equation \cite{Torquatobook, Coniglio1977, Bug1986}
		\begin{align}
			P(\br, \bu, \bu')=&C^+(\br, \bu, \bu') \label{cOZE}\\
				&+n \int \dd \br' \langle C^+(\br', \bu, \bu'') P(\br-\br', \bu'', \bu') \rangle''. \nonumber
		\end{align}
		Here, the direct connectedness function $C^+(\br, \bu, \bu')$ measures the probability that two test particles are part of the same cluster, and that they cannot become disconnected from each other upon the removal of any other single particle from the cluster.
		It is, in fact, the ``connectedness'' part of the direct correlation function $C(\br, \bu, \bu') = C^+(\br, \bu, \bu') + C^*(\br, \bu, \bu')$, with  $C^*(\br, \bu, \bu')$  the ``blocking'' part.
		The convolution term on the right-hand-side of Eq.\,\eqref{cOZE} describes all other particle clusters that contain at least one ``bottleneck'' particle, which, upon removal, leaves the two test particles disconnected.
		It is important to note that the particle density $n$ enters Eq. \eqref{cOZE} directly, but also indirectly through the orientational averaging, as the orientational distribution $\psi (\bu)$ depends on the density via Eq. \eqref{Onsager}.
		
		To investigate percolation, it turns out useful to take the Fourier transform of the connectedness Ornstein-Zernike equation, and to focus on clusters at the macroscopic scale by taking the limit of vanishing wave vectors. 
		This is equivalent to the volume integral $\hat{(\cdots)} = \int \dd \br (\cdots)$ of Eq. \eqref{cOZE}, and yields
		\begin{align}
			\hat{P}(\bu, \bu')=\hat{C}^+(\bu, \bu')+n \langle \hat{C}^+(\bu, \bu'') \hat{P}(\bu'', \bu') \rangle''. \label{FTcOZE}
		\end{align}
		The weight-average number of particles within a cluster is then given by the average cluster size, \cite{Torquatobook, Coniglio1977, Bug1986}
		\begin{equation}
			S = 1 + n \langle \langle \hat{P}(\bu, \bu') \rangle \rangle'. \label{S}
		\end{equation}
		At percolation, particle clusters grow infinitely large, and the cluster size diverges.
		In order to make quantitative predictions for the cluster size $S$ and establish the percolation threshold, we need a closure relation for Eq.\,\eqref{FTcOZE}, that is, a reasonable estimate for $\hat{C}^+(\bu, \bu')$.

		It is known that the second virial approximation,  $\hat{C}^+(\bu,\bu') = \hat{f}^+(\bu, \bu')$ provides an accurate percolation closure for particles with aspect ratios above $\sim300$. \cite{Otten2011, Schilling2015}
		Here, $\hat{f}^+(\bu, \bu')$ is the volume integral of the connectedness Mayer function $f^+(\br, \bu, \bu') = \exp[ -\beta U^+(\br, \bu, \bu')]$, with $\beta$ the reciprocal thermal energy. \cite{Otten2011, Schilling2015}
		The connectedness potential $U^+$ is infinite for pairs of particle that are not directly connected, and zero for particles that are connected, implying that $f^+ = 1$ if particles are connected, and $f^+ = 0$ otherwise.
		Disconnected configurations include forbidden configurations, where the hard cores of the particles overlap, and those for which the shortest surface-to-surface distance is larger than the connectivity criterion $\lambda$.
		As a result, the volume integral of the connectedness Mayer function of straight spherocylinders is equivalent to their contact volume,\,\cite{Coniglio1977, Onsager1949, Balberg1984, Balberg1984_2}
		\begin{align}
			\hat{f}^+(\bu,\bu') = & 2L^2 \lambda |\bu \times \bu'| + 2\pi L \left[ (D+\lambda)^2 - D^2\right] \nonumber \\
			&\hspace{1.95cm} +\frac{4\pi}{3}\left[ (D+\lambda)^3 - D^3\right], \label{f0}
		\end{align}
		\textit{i.e.}, the volume in which one particle with orientation $\bu$ can be located such that its contact shell overlaps with that of a second, fixed test particle of orientation $\bu'$.

		To investigate percolation of nanoparticles with aspect ratios smaller than $300$, we must go beyond the second virial approximation.
		It seems straightforward to effectively account for higher-order virial contributions by renormalizing the \textit{connectedness} Mayer function, $f^+$, in a similar way as the Mayer function $f$ in the context of the Onsager theory.
		For the Lee-Parsons approach, this straightforwardly translates into $\hat{C}^+(\bu, \bu') = \Gamma_\text{LP}(\phi) \hat{f}^+(\bu, \bu')$.  \cite{Schilling2015}
		For the Scaled Particle Theory approach, we insert $f(\br, \bu, \bu') = f^+(\br, \bu, \bu') + f^*(\br, \bu, \bu')$ into the direct correlation function and obtain
		\begin{align}
			C(\br, \bu, \bu') =& \Gamma_\text{SPT}(\phi, L/D) f(\br, \bu, \bu') + \Omega_\text{SPT}(\br, n, L, D) \nonumber\\
					=& \Gamma_\text{SPT}(\phi, L/D) f^+(\br, \bu, \bu') \\
						&+ \Gamma_\text{SPT}(\phi, L/D) f^*(\br, \bu, \bu') + \Omega_\text{SPT}(\br, n, L, D). \nonumber
		\end{align}
		Here, $f^*(\br, \bu, \bu')$ is the blocking Mayer function.
		
		At this point, it is not straightforward to decide if the orientation-independent term $\Omega_\text{SPT}(\br, n, L, D)$ should contribute to the connectedness or the blocking part of the direct correlation function.
		For reasons of simplicity, and by analogy to the Lee-Parsons closure, we choose to contract it into the blocking part, $\hat{C}^*(\bu, \bu') = \Gamma_\text{SPT}(\phi, L/D) \hat{f}^*(\bu, \bu')  + \hat{\Omega}_\text{SPT}(n, L, D)$, so that our connectedness closure within Scaled Particle Theory reads
		\begin{align}
			\hat{C}^+(\bu, \bu') = \Gamma_\text{SPT}(\phi, L/D) \hat{f}^+(\bu, \bu').
		\end{align}
		Given the scaling factors $\Gamma_\text{LP}(\phi)$ and $\Gamma_\text{SPT}(\phi, L/D)$, we can now obtain quantitative predictions for the percolation threshold by solving Eqs. \eqref{Onsager} and \eqref{S} numerically, but also analytically by invoking a Schwinger-type variational theory, which we outline in the following Section.

	\section{Variational percolation theory \label{sec:vartheory}}
		In order to set up our variational theory,\,\cite{Finner2019PRL} first applied by Odijk in the context of radiation scattering from concentrated solutions of hard rods, \cite{vdSchoot1990} it turns out useful to rewrite the governing percolation equations by defining the function
		\begin{align}
			m(\bu) = \sqrt{\psi(\bu)}\left[1+n\langle \hat{P}(\bu, \bu') \rangle' \right]. \label{mDef}
		\end{align}
		With Eq.~\eqref{mDef}, the average cluster size can be written as:
		\begin{equation}
			S=\int \dd \bu \sqrt{\psi(\bu)} m(\bu),
		\end{equation}
		and the connectedness Ornstein-Zernike equation, averaged over one angular degree of freedom, reads
		\begin{equation}
			m(\bu)=\sqrt{\psi(\bu)} + n \int \dd\bu' K(\bu, \bu') m(\bu'). \label{rcOZE}
		\end{equation}
		Here,
		\begin{equation}
			K(\bu, \bu') = \sqrt{\psi(\bu)} \hat{C}^+(\bu, \bu') \sqrt{\psi(\bu')} \label{kernelK}
		\end{equation}
		denotes the kernel associated with the integral operator applied to the function $m(\bu)$.
		Note that this kernel is symmetric in $\bu$ and $\bu'$, as $\hat{C}^+(\bu, \bu')=\hat{C}^+(\bu',\bu)$ due to particle exchange symmetry. \cite{Perera1999}

		Now consider the functional
		\begin{align}
			F[m] =& \int \dd\bu \left[ \frac{m^2(\bu)}{2} - \sqrt{\psi(\bu)}  m(\bu) \right] \label{functional}\\
				& -\frac{n}{2} \int \dd\bu  \int \dd \bu' m(\bu) K(\bu, \bu') m(\bu'). \nonumber
		\end{align}
		It is straightforward to show that the function $m(\bu)$ that functionally extremizes Eq. \eqref{functional}, obeys the reduced connectedness Ornstein-Zernike equation given by Eq.\,\eqref{rcOZE}.
		Moreover, we notice that, in the limit of $n \rightarrow 0$, the solution to Eq. \eqref{rcOZE} obeys $m(\bu)=\sqrt{\psi(\bu)}$.
		Hence, it seems sensible to choose $m(\bu)= M \sqrt{\psi(\bu)}$ as a plausible trial function, with $M$ a variational parameter.
		Inserting $m(\bu)$ into Eq. \eqref{functional}, and calculating the stationary value of $M$ by setting $\partial F / \partial M =0$, we find $M = \big(1-n \langle \langle \hat{C}^+(\bu, \bu') \rangle \rangle'\big)^{-1}$.
		It follows that the cluster size
		\begin{align}
			S^{-1} = 1-n \langle \langle \hat{C}^+(\bu, \bu') \rangle \rangle' \label{varS}
		\end{align}
		 diverges under the condition
		\begin{equation}
			n \langle \langle \hat{C}^+(\bu, \bu') \rangle \rangle' = 1. \label{varPT}
		\end{equation}
		
		For isotropic solutions and for perfectly (uniaxially) aligned particles, the analytical predictions of \eqref{varS} and \eqref{varPT} are exact. \cite{Drwenski2017}
		In the isotropic phase,  they also show very good agreement with Monte Carlo simulations, given an appropriate choice of the renormalization factor $\Gamma(\phi, L/D)$.\,\cite{Schilling2015}
		Deviations of Eq.\,\eqref{varPT} from the (exact) numerical solution do occur if particles are partially aligned due to an external orienting field, \textit{i.e.}, in paranematic dispersions.\,\cite{Otten2012, Finner2018}
		Comparison with our numerical solution shows that Eqs. \eqref{varS} and \eqref{varPT} also provide an excellent estimate for the generalized cluster size and the percolation threshold in the uniaxial nematic phase, as we shall see later in this article.
		
		In principle, the accuracy of our analytical predictions could be further improved by using a more sophisticated variational trial function, for instance of the form $m(\bu)=\sqrt{\psi(\bu)}\big(M+N \theta^2\big)$.
		Here, $M$ and $N$ are variational parameters, the stationary values of which are obtained by setting $\partial F / \partial M = \partial F / \partial N =0$.
		Note that using this more sophisticated trial function only leads to small quantitative changes in the percolation threshold and does not affect our analytical predictions qualitatively.
		We therefore decide to use the one-parameter trial function and postpone a discussion of the choice of trial function to a later point in this article.

		To make our analytical predictions on percolation in the uniaxial nematic quantitative, we insert our closure $\hat{C}^+(\bu, \bu') = \Gamma(\phi, L/D) \hat{f}^+( \bu, \bu')$ into Eq.\,\eqref{varPT} and calculate the resulting orientational averages by invoking Odijk's Gaussian approximation.
		With $\langle \langle |\bu \times \bu'| \rangle \rangle' \sim \pi^{1/2} \alpha^{-1/2}$, and $\langle \langle \theta^2 |\bu \times \bu'| \rangle \rangle' \sim 5 \pi^{1/2} \alpha^{-3/2}/2$\,\cite{Vroege1992}, the resulting inverse cluster size reads
		\begin{equation}
			S^{-1} = 1 - 4 \frac{\lambda}{D} -  \phi \Gamma\left(\phi, \frac{L}{D}\right) h\left(\frac{\lambda}{D}, \frac{L}{D}\right).
		\end{equation}
		Here,
		\begin{equation}
			h\left(\frac{\lambda}{D}, \frac{L}{D}\right)=\frac{8\left[\left(1+\frac{\lambda}{D}\right)^2-1\right]+\frac{4\pi}{3}\frac{D}{L}\left[\left(1+\frac{\lambda}{D}\right)^3-1\right]}{\left[1+\frac{2}{3}\frac{D}{L}\right]} \label{h}
		\end{equation}
		is a function of the relative contact shell thickness $\lambda/D$ of the spherocylinders and the particle aspect ratio $L/D$.
		We conclude that percolation occurs if
		\begin{equation}
			4\frac{\lambda}{D}+  \phi \Gamma\left(\phi, \frac{L}{D}\right) h\left(\frac{\lambda}{D}, \frac{L}{D}\right) \geq 1, \label{varPercLP}
		\end{equation}
		with $\Gamma(\phi, L/D) \geq 1$.
		Eq.\,\eqref{varPercLP} implies that, irrespective of how short the rods are (provided that they do support a stable nematic phase), percolation in the nematic phase occurs \textit{always} if $\lambda \geq D/4$, consistent with recently published results for infinitely slender nanoparticles. \cite{Finner2019PRL}
		What is different for smaller aspect ratios is that percolation may also be found for $\lambda < D/4$, provided that the volume fraction $\phi$ is sufficiently large.
		How large, depends on the aspect ratio $L/D$, on the contact shell thickness $\lambda/D$, and on the particular scaling factor $\Gamma(\phi, L/D)$, for which we provide two alternative expressions in the following Section.

	%%%%         CLOSURE      %%%%
	\section{Scaling factors \label{sec:gammas}}

		One reasonable way of rescaling the Mayer function and the connectedness Mayer function of spherocylinders to account for higher-order virial contributions is based on the theory of Parsons and of Lee, which makes use of an interpolation between the Carnahan-Starling equation of state of hard spheres and the second virial equation of state of infinitely slender hard rods.\,\cite{Parsons1979, Lee1987}
		Within Lee-Parsons theory, the excluded volume of hard spherocylinders is, by construction, rescaled by the factor
		\begin{align}
			\Gamma_\text{LP}(\phi) = \frac{1- 3 \phi /4}{(1-\phi)^2}, \label{GammaLP}
		\end{align}
		and the resulting orientational distribution function obeys Eq.\,\eqref{Onsager} with $\Gamma(\phi, L/D) \rightarrow \Gamma_\text{LP}(\phi)$.
		This correction to the second virial approximation has been shown to provide very good predictions for the isotropic-to-nematic transition densities of hard spherocylinders.  \cite{Parsons1979, Lee1987, Bolhuis1997}
		In fact, the connectedness closure $\hat{C}^+ (\bu, \bu') = \Gamma_\text{LP}(\phi) \hat{f}^+(\bu, \bu')$ also gives highly accurate results for the percolation threshold, at least in the isotropic phase and for aspect ratios above $\sim10$, as has recently been analysed in a comparison between connectedness percolation theory and Monte Carlo simulations. \cite{Schilling2015, Meyer2015}
		Our hope is that the Lee-Parsons closure also provides accurate predictions for the percolation threshold in the uniaxial nematic phase.%, which we will be discussing in Section \ref{sec:results} of this article.

		A second way of deriving a correction to the second virial approximation is by invoking Scaled Particle Theory:
		a framework which makes use of an interpolation between the reversible work needed to insert a very \textit{small} particle and a very \textit{large} particle into the suspension. \cite{TuinierBook, Cotter1977, Reiss1959}
		Within this framework, the volume integral of the direct correlation function is given by $\hat{C}(\bu, \bu') = \Gamma_\text{SPT}(\phi, L/D) \hat{f}(\bu, \bu') + \hat{\Omega}_\text{SPT}(n, L, D)$, with the scaling factor
		\begin{align}
			\Gamma_\text{SPT}( \phi, L / D ) = \frac{1}{1-\phi} \Big[ 1 + \frac{2+ 2 L / D}{2 + 3 L / D} \frac{\phi}{1-\phi} \Big], \label{GammaSPT}
		\end{align}
		and
		\begin{align}
			\hat{\Omega}_\text{SPT}(\phi, L, D) =&\frac{\pi D^2 (3L+2D)}{6(1-\phi)}\Big[ 1 + \frac{(1-\phi)\ln(1-\phi)}{\phi}\\
			 & + \frac{\phi}{1-\phi} \frac{2 + 2 L / D}{2 + 3L / D} \left(4 - 3\frac{1 + 2 L / D}{2 + 3 L / D}\right) \Big] \nonumber
		\end{align}
		an additional correction term.
		The distribution function of particle orientations obeys Eq. \eqref{Onsager} with $\Gamma(\phi, L/D) \rightarrow \Gamma_\text{SPT}(\phi, L/D)$.
		For the derivation of the corrections $\Gamma_\text{SPT}(\phi, L/D)$ and $\hat{\Omega}_\text{SPT}(n, L, D)$, we refer the reader to Appendix \ref{app:A}.
		As discussed in Section \ref{sec:percTheory}, when applying Scaled Particle Theory to the percolation problem, we have the freedom to choose if the correction term $\Omega_\text{SPT}(\br, n, L, D)$ impacts upon the direct connectedness function, $C^+(\br, \bu, \bu')$, or the direct blocking function, $C^*(\br, \bu, \bu')$.
		For simplicity, we choose to contract it into the blocking function and obtain the connectedness closure $\hat{C}^+ (\bu, \bu') = \Gamma_\text{SPT}(\phi) \hat{f}^+(\bu, \bu')$.

		The question is now: which of the two approaches is more accurate at predicting the percolation threshold of hard spherocylinders of finite aspect ratio, and does the answer depend on the symmetry of the underlying phase?
		What is known is that Scaled Particle Theory reproduces the third virial coefficient more accurately than Lee-Parsons theory does. \cite{Vroege1992}
		However, the isotropic-nematic coexistence densities obtained by either approach are roughly equally accurate when compared to results of Monte Carlo simulations, with Lee-Parsons theory having a slight advantage for aspect ratios below $\sim25$ (see also Figure \ref{fig:coexistence}). \cite{Vroege1992, TuinierBook, Bolhuis1997}
		In Section \ref{sec:results}, we compare the accuracy of our two percolation closures across the isotropic-nematic phase transition.
		For this purpose, we calculate the percolation threshold for various aspect ratios both numerically and analytically with our variational theory, and compare the results to our corresponding Monte Carlo simulations.
		Before presenting our results, however, let us first describe the details of our simulations and our numerical procedures in the following Section.% \ref{sec:numericsMC}.

%%%%%%%%%%%%%%%%%%%%%%%%%%%

	\section{Numerical details and Monte Carlo simulations\label{sec:numericsMC}}
		To determine the percolation threshold numerically, we first calculate the orientational distribution function, $\psi(\theta)$, in the nematic phase for a fixed value of the particle density $n$ (or volume fraction $\phi$).
		We do this by solving the generalized Onsager equation, Eq.\,\eqref{Onsager}, using a recursive iteration scheme, as described in Ref.\,\cite{vanRoij2005}.
		The iteration is performed on an angular grid of $N_\theta = 400$ and $N_\phi = 400$ grid points, with polar angles $0 \leq \theta \leq \pi/2$ and azimuthal angles $0 \leq \phi \leq 2 \pi$.
		To increase the resolution around the peak of the orientational distribution function, we divide our $\theta$-grid into three equidistant grids with $N_\theta/2$ points in the range $[0,\pi/8)$, $N_\theta/4$ points in $[\pi/8,\pi / 4)$ and  $N_\theta/4$ grid points in $[\pi/4,\pi/2)$.
		Using the Gaussian distribution, $\psi(\theta)=c^2 \Gamma^2(\phi, L/D) \exp (- 2c^2 \Gamma^2(\phi, L/D)  \theta^2/ \pi) / \pi^2$ for $0 \leq \theta \leq \pi / 2$, and $\psi(\pi - \theta)$ for $\pi/2 \leq \theta \leq \pi$, as an initial guess, we iterate Eq.\,\eqref{Onsager} until the difference between subsequent iterations of $\psi(\theta)$ at each grid point is smaller than our iteration tolerance of $10^{-8}$.
		The volume fractions at phase coexistence in the isotropic ($\phi_\mathrm{iso}$) and in the nematic phase ($\phi_\mathrm{nem}$), are calculated by equating the pressures $p(\phi_\mathrm{iso}) = p(\phi_\mathrm{nem})$ and the chemical potentials $\mu(\phi_\mathrm{iso}) = \mu(\phi_\mathrm{nem})$ in the two phases, as described in Ref. \cite{vanRoij2005}.
		Between the coexistence densities, our solutions for the orientational distribution function are metastable, and are therefore not accessible in thermodynamic equilibrium.

		With the numerically exact solution for the orientational distribution, $\psi(\theta)$, we calculate the percolation threshold in terms of the critical contact shell thickness $\lambda_\pp / D$ for the same (fixed) particle density, $n$.
		This we do by averaging the connectedness Ornstein-Zernike equation \eqref{cOZE} over one orientation $\bu'$ and inserting our closure relation $\hat{C}^+(\bu, \bu') = \Gamma(\phi, L/D) \hat{f}^+(\bu, \bu')$.
		Subsequently, we perform a sweep in the connectedness criterion $\lambda / D$, with steps of $\Delta_\lambda = 0.002$.
		For each $\lambda / D$, we obtain a discrete representation of the function $\langle \hat{P}(\bu, \bu')\rangle'$ by recursive iteration.
		Our iteration scheme is analogous to that of the generalized Onsager equation described above, and with the same angular grid and iteration tolerance.
		As an initial guess, we choose the source term $\langle \hat{P}(\bu, \bu')\rangle' =  \langle \hat{C}^+(\bu, \bu') \rangle' $.

		Below the critical shell thickness $\lambda_\pp / D$, our iteration converges, and we calculate the inverse of the average cluster size, $S^{-1}$.
		If the iteration does not converge, and the difference between subsequent iterations grows within the first 1000 iteration steps, we abort the iteration and assume a percolating network.
		This procedure provides an estimate of the critical shell thickness within an interval of  $ \pm \Delta_\lambda / 2$.
		For a more accurate prediction of the percolation threshold, we use linear extrapolation of the last two data points of $S^{-1}(\lambda / D)$ below percolation and determine the root of the resulting linear function.
		This scheme is repeated for all particle densities $n$.\\

		In the Monte Carlo (MC) simulations, we initialize particle configurations either on a regular grid with perfect orientational alignment, or according to a uniform and random distribution of positions and orientations. In the latter case we need an additional initialization procedure to exclude overlaps. We first use a soft rod-rod interaction potential
		\begin{equation}\label{eq:SoftPotential}
			V(r_{ij})=
	       		\begin{cases}
	        		a<\infty, & \text{if}\ r_{ij}<D \\
	       			0, & \text{otherwise,}
	       		\end{cases}
	    	\end{equation}
		so overlaps between rods are initially allowed with a nonzero probability proportional to the Boltzmann weight $\exp-\beta V(r_{ij})$.
	    	Here, $r_{ij}$ denotes the shortest distance between the long axes of two rods $i$ and $j$. 
	    	The initialization is finalized by running short MC simulations, where the overlap cost $a$ is gradually increased until the system is devoid of overlaps.
	    	Equilibrium configurations are then generated using canonical MC simulations in a cuboid simulation box with periodic boundary conditions. \cite{Miller2009}
	    	The equilibration is monitored using the mean acceptance probabilities of the particle displacement and rotation moves, together with the nematic order parameter $S_2= \langle 3 \cos^2 \theta - 1 \rangle  /2 $, distinguishing between the isotropic phase ($|S_2| < 0.05$, corresponding to uniform distribution of orientations over all solid angles) and the nematic phase ($S_2 > 0.5$).
	    	
	    	To estimate the percolation threshold, $\lambda_\pp / D$, for a given aspect ratio and system size, we generate an ensemble of $200-5000$ independent equilibrium configurations at a fixed volume fraction $\phi = N v_\text{core}/V,$ with $N$ the number of rods, $V$ the volume of our simulation box, and $v_\text{core}=\pi D^2 (L/4+D/6)$ the hard-core volume of one rod.
		Subsequently, we perform a sweep in the shell thickness $\lambda /D$, in which we calculate the percolation probability, $p(\lambda)$, as the fraction of configurations containing a percolating cluster.
	    	A cluster is deemed percolating, if each constituent rod is connected to its own periodic image in at least one periodic direction of the simulation box.\,\cite{Skvor07a}
	    	As a consequence of finite system sizes, the so-obtained percolation probability curves are well described by the hyperbolic tangent function, which becomes narrower for larger system sizes. 
	    	We carried out simulations of suspensions of monodisperse rods with aspect ratios $L/D=5,10,20,50$ and $100$.
		For each aspect ratio and each simulation box size, the  volume fractions spanned a range that covered both the isotropic and the nematic phase, see Table~\ref{tab:simulparams}. 
	
		\begin{table}
			\begin{center}
				\begin{tabular}{| c | c | c |}
					\hline
					$L/D$ &\ Box dimensions $[\mathcal{L}_{x} = \mathcal{L}_y; \mathcal{L}_z ]$ & $\phi$ \\ \hline
					$5$  & $[5L;5L]$ & $[0.25\text{ to }0.54]$ \\ \hline
					$10$ & $[4L \text{ to } 6L; 4L \text{ to } 6L]$ & $[0.19 \text{ to } 0.30]$ \\ \hline
					$20$  & $[3L\text{ to } 6L; 4L \text{ to } 8L]$ & $[0.065 \text{ to } 0.282]$ \\ \hline
					$50$  & $[3L \text{ to } 4L; 3L \text{ to } 6L]$ & $[0.025 \text{ to } 0.085]$ \\ \hline
					$100$ & $[2.2L \text{ to } 6L; 2.2L \text{ to } 8L]$ & $[0.02 \text{ to } 0.08]$ \\ \hline
				\end{tabular}
			\end{center}
			\caption{MC simulation parameters for suspension of hard rods. 
			From left to right the columns show: the particle aspect ratio $L/D$, the range of simulated box dimensions $\mathcal{L}_{x}, \mathcal{L}_y$ and $\mathcal{L}_z$  in units of the rod length $L$, and the range of simulated volume fractions $\phi$.}
			\label{tab:simulparams}
                      \end{table}
                      
		In the isotropic phase, and for the box sizes used, the volume fraction at which percolation sets in is nearly independent of the system size; that is, there are finite-size effects, but these are small compared to the impact of $\lambda$ and $\phi$ on the percolation probability. \cite{StaufferBook}
		 For the purpose of this article, which lies in analyzing the trend of $\lambda_\pp(\phi),$ it is therefore sufficient to sample the $p(\lambda)$-curve for one box size per volume fraction, and to use the point $p(\lambda_\pp)=0.99$ as an estimate of the percolation threshold $\lambda_\pp$. \footnote{For $L/D=100$, we used $p(\lambda_\pp)=0.5$ instead of $p(\lambda_\pp)=0.99$. As the systems were large enough for $p(\lambda)$ to be nearly a step function, the effect on the threshold value was smaller than $1\%$.}
		In the isotropic and paranematic phase, $p(\lambda)$-curves for different system sizes cross each other in one point, which denotes the percolation threshold in the thermodynamic limit. \cite{Skvor07a, Finner2018} This is a direct consequence of the scale invariance of percolation on approach of the critical point, which can be estimated as described without making any assumptions on critical exponents. However, in the nematic phase, a finite-size scaling analysis of the percolation transition becomes more intricate. 
		This is because in the nematic, the position of the crossing point of $p(\lambda)$-curves depends on the aspect ratio of the box relative to the aspect ratio of particle clusters, as clusters change in size as well as in shape on approach of the percolation transition.
		A rigorous finite-size scaling analysis in the nematic phase is therefore non-trivial: it needs to be performed with boxes of varying sizes \textit{and} aspect ratios.
		
		For $L/D=100$, we study boxes of aspect ratios $\mathcal{L}_z / \mathcal{L}_x$ ranging from  $2/3$ to $8$, and find that above a box aspect ratio of $3$ the systematic error due to finite-size effects and box-shape effects is negligible compared to the variations in $\lambda_\pp$ caused by changes in the volume fraction. 
		Thus, the comparison to the theoretical curves presented here is not affected by finite-size effects.
		
		Next, we briefly explain how the coexistence densities are determined in our MC-simulations.
		For aspect ratios between $10$ and $50$, a highly elongated simulation box is chosen for a volume fraction of the suspension within the coexistence region.
		During the equilibration phase of the suspension, given the highly anisotropic shape of the box (e.g., $\mathcal{L}_y = \mathcal{L}_z=3L$ and $\mathcal{L}_x=15L$ for $L/D=20$), the rods are able to undergo phase separation into distinct slabs of isotropic and nematic phase, as this minimizes the contact area between the two coexisting phases (reducing it to the smallest cross section of the box). 
		By initializing the suspension in the nematic phase, the rods are less likely to become jammed in the separation process.
		
		We compute the density profile along the elongated dimension. 
		In this profile, the formed slabs of each phase appear as regions of constant density with a negligible spread (given sufficient averaging of the profile). 
		Finally, by fitting a constant function to said regions, the binodal densities are estimated.
		Note that for the data points near (or within) the so-obtained coexistence windows, determining if the mixture is in fact stable or metastable would only be feasible by computing and comparing the respective chemical potentials in the isotropic and nematic phase.
		However, due to CPU time limitations, and as we have not observed any instabilities during the simulation of the equilibrated suspensions, we do not perform additional free energy calculations to this end.

%%%%%%%%%%%%%%%%%%%%%%%%%%%

	\section{Results for percolation in the nematic phase \label{sec:results}}
\begin{figure*}
			\includegraphics[width = \textwidth]{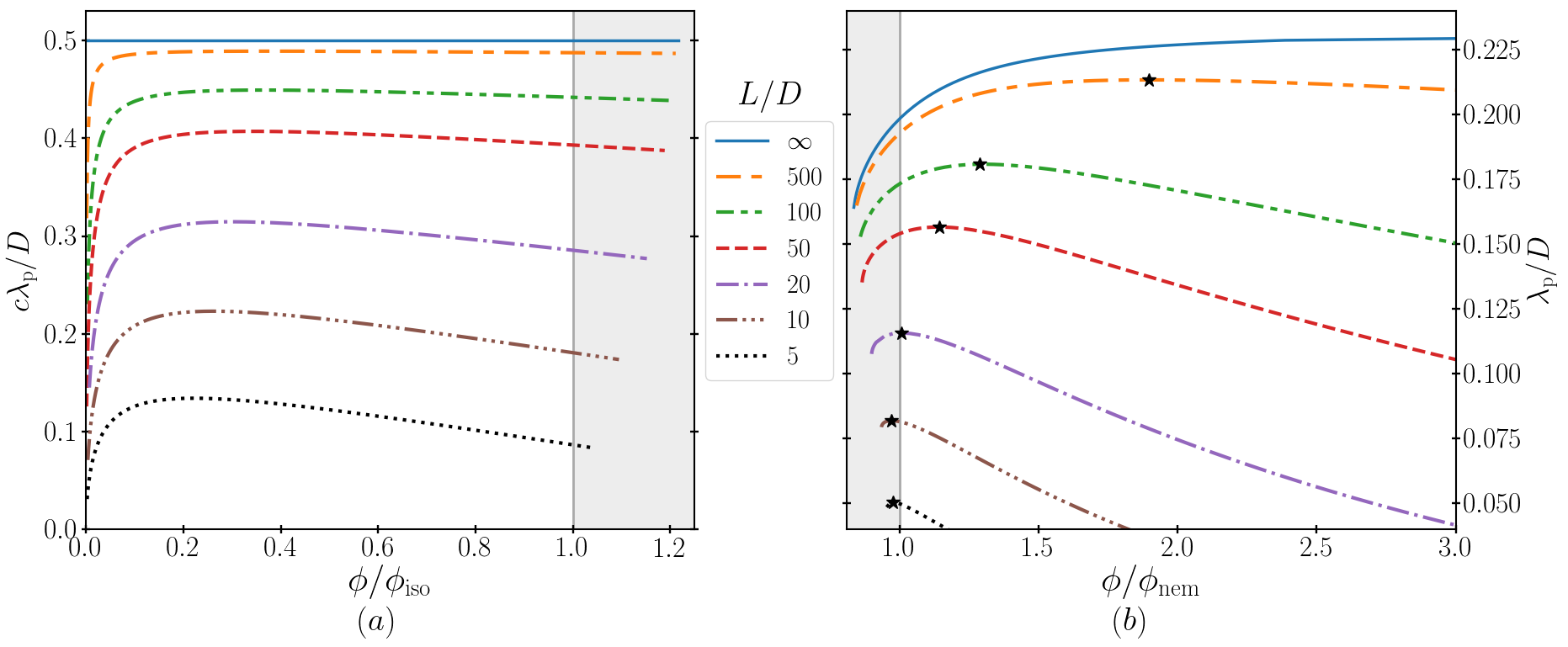}
			\caption{
				Percolation thresholds of hard spherocylinders with aspect ratios between five and infinity in the isotropic (a) and in the nematic (b) phase.
				On the horizontal axes, the volume fractions are scaled to the isotropic and nematic binodals, $\phi_\mathrm{iso}$ and $\phi_\mathrm{nem}$, respectively.
				For convenience, the percolation threshold in Figure \ref{fig:scaledPlot} (a) is rescaled with the dimensionless particle concentration $c = \phi (D/L+2D^2/3L^2)^{-1}$.
				The blue solid lines show the known theoretical predictions in the Onsager limit $L/D\rightarrow \infty$. \cite{Otten2011, Finner2019PRL}
				In the coexistence region, denoted by the grey shaded areas, our solutions are metastable, and the curves end at the isotropic (a) and nematic (b) spinodals, \textit{i.e.}, the filler fractions at which the respective orientational distribution function, $\psi(\theta)$, becomes unstable. \cite{Odijk1986,vanRoij2005}
				The black stars in Figure \ref{fig:scaledPlot} (b) mark the local maxima of the percolation curves in the nematic phase.
			}
			\label{fig:scaledPlot}
		\end{figure*}

	%%% GENERAL %%%
		For spherocylinders in the Onsager limit $L/D\rightarrow \infty$, we have recently shown theoretically that percolation in the nematic phase can be lost upon the addition of particles, and that the cluster size deep in the nematic is virtually independent of the particle density. \cite{Finner2019PRL}
		In order to investigate how these predictions change for nanoparticles with a \textit{finite} aspect ratio, we compare in Figure \ref{fig:scaledPlot} our theoretical predictons for spherocylinders of aspect ratios between five and infinity.
		
\begin{figure}
			\includegraphics[width = 0.95 \linewidth]{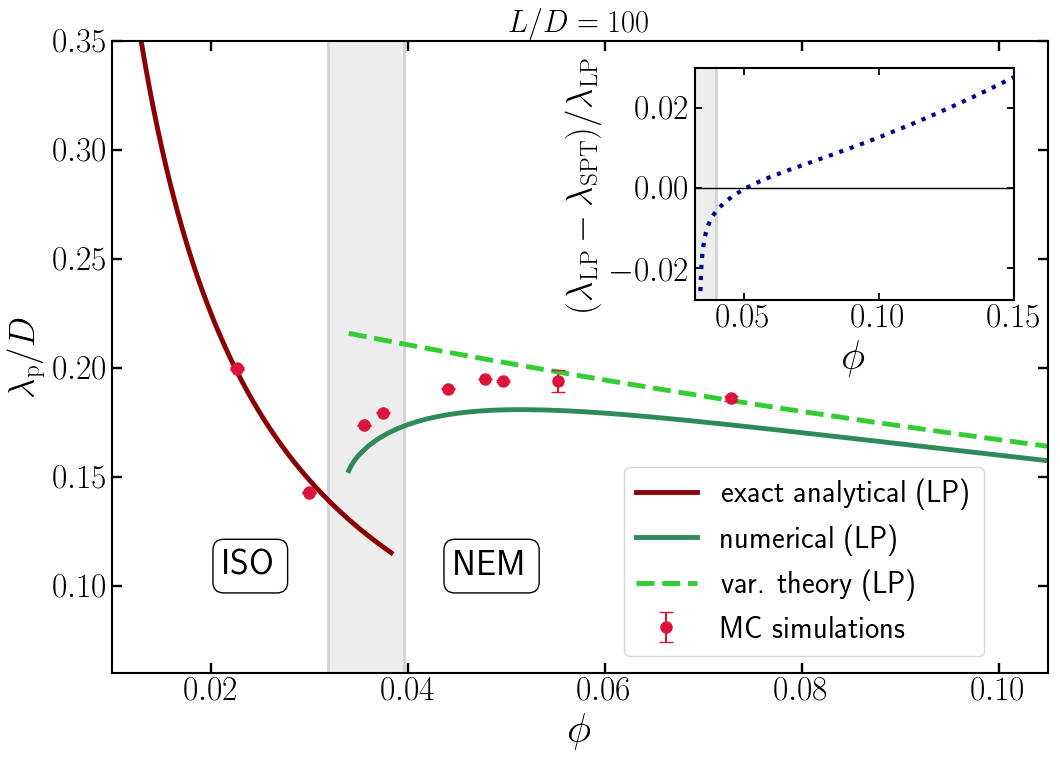}
			\caption{
				Critical shell thickness $\lambda_\pp/D$ at percolation as a function of the filler fraction $\phi$ for a particle aspect ratio $L/D=100$.
				The solid lines represent our (numerically) exact results in the isotropic phase (brown) and in the uniaxial nematic (dark green), obtained using Lee-Parsons theory and the Lee-Parsons closure relation (LP).
				The light-green dashed line indicates the result of our variational theory with the Lee-Parsons closure.
				Monte Carlo simulation data are re-plotted from Ref. \cite{Finner2019PRL} and indicated by red circles.
				Also indicated is the isotropic-nematic coexistence region according to Lee-Parsons theory (grey shaded area).
				The dispersions with volume fractions within this region appear stable in our MC simulations, but might actually be metastable, see Section \ref{sec:numericsMC}.
				Inset: relative difference of the percolation thresholds obtained with the Lee-Parsons approach, Eq. \eqref{GammaLP}, and the Scaled Particle Theory approach, Eq. \eqref{GammaSPT}.
			}
			\label{fig:PD100}
		\end{figure}

		Figure \ref{fig:scaledPlot} (a) shows the critical connectedness range $\lambda_\pp/D$ in the isotropic phase, as a function of the volume fraction $\phi$ relative to the isotropic binodal $\phi_\text{iso}$.
		The grey shaded area visualizes the coexistence region.
		Our results are obtained with Lee-Parsons theory in combination with the Lee-Parsons closure for percolation, Eq. \eqref{GammaLP}. \cite{Schilling2015}
		For convenience, we rescale the vertical axis with the particle concentration, $c \equiv \phi (D/L+2D^2/3L^2)^{-1}$.
		The Figure demonstrates the deviation in the slope of the percolation curve from  $\lambda_\pp / D = (2c)^{-1}$, which is the exact result in the limit of infinite aspect ratios. \cite{Otten2012}
		For short rods, the isotropic binodal, $\phi_\text{iso}$, is located at high filler fractions, so that higher-order virial terms come into play and aid the formation of a system-spanning cluster.
		This explains why the scaled percolation thresholds in Figure \ref{fig:scaledPlot} (a) decrease with decreasing particle aspect ratio.
		Note that the apparent maximum in the percolation curves is due to the scaling of the axes, and that the percolation threshold $\lambda_\pp / D$ in the isotropic phase is a strictly monotonically decreasing function of the filler fraction, as can be seen, for instance, in Figure \ref{fig:PD100}.

		Figure \ref{fig:scaledPlot} (b) shows our numerical predictions for the critical connectedness shell $\lambda_\pp / D$ in the \textit{nematic} phase as a function of the volume fraction scaled to the nematic binodal, $\phi_\text{nem}$.
		Again, our predictions are obtained with the Lee-Parsons closure, Eq. \eqref{GammaLP}, and the grey area denotes the coexistence region.
		We find that percolation may be lost upon entering the nematic phase, and upon the addition of particles within the nematic phase, if the connectedness range is sufficiently small (see also Figure \ref{fig:PD100}).
		Above a critical connectedness range, there is \textit{always} percolation in the nematic, regardless of the filler fraction.
		These findings are in agreement with our earlier results for infinitely slender particles in the second virial approximation (the blue solid line in Figure \ref{fig:scaledPlot} (b)). \cite{Finner2019PRL}
		The difference with the case of \textit{finite} aspect ratios is that the shell thickness above which we find percolation is lower than that in the Onsager limit, and that the percolation threshold $\lambda_\pp / D$ exhibits a local maximum instead of saturating at large filler fractions.
		The latter means that percolation in the nematic cannot only be lost upon the addition of filler particles, it can also be re-gained if the particle density is increased even further: percolation in the nematic exhibits re-entrance behavior.
		At large enough volume fractions, percolation in the uniaxial nematic may therefore occur even if the contact shell is too thin to allow for percolation in the \textit{isotropic} phase (see also, e.g., Figure \ref{fig:PD50}).
		 % 0.14, for L/D=20 at 0.21 | for L/D=10 at 0.28/0.29)

		%%% 100 %%%
		
		A more subtle effect that arises for rods of decreasing aspect ratios is that the maximum of the percolation curve in the nematic phase, denoted by the black star symbols in Figure \ref{fig:scaledPlot} (b), shifts to lower volume fractions.
		In fact, for aspect ratios below $L/D \simeq 20$, the local maximum lies \textit{within} the coexistence region and is therefore not accessible in thermodynamically equilibrated, macroscopic suspensions.
		This means that, if the aspect ratio is sufficiently small, the percolation threshold in the stable nematic (in terms of the critical shell thickness) becomes a monotonically decreasing function of the filler fraction.
		As a result, we can no longer \textit{lose} percolation by adding more particles to the suspension, only \textit{gain} it -- a behavior that is well known from the isotropic phase and contrary to that in the slender rod limit. \cite{Finner2019PRL}
		Notice also that the percolation threshold near the melting transition to the isotropic phase is only modestly sensitive to the aspect ratio, which is unusual because the isotropic-to-nematic transition density itself is strongly dependent on the particle aspect ratio.\cite{Odijk1986}

		Unlike the numerical results, our \textit{analytical} solution for the percolation threshold in the nematic, indicated in Figure \ref{fig:PD100} by the dashed light green line, does not exhibit a local maximum in $\phi$.
		Close to the I-N transition, it decreases monotonically as $1/4 - 9 c D/8L$, at least in the limit of large aspect ratios $L/D$.
		This is shown in Figure \ref{fig:PD100}, where we plot both the analytical and the numerical predictions for the percolation threshold of rods with aspect ratio $L / D = 100$.
		The monotonicity of our analytical solution is a direct consequence of the Gaussian approximation to the orientational distribution function.
		Interestingly, for rods of finite aspect ratio, the analytical solution with the one-parameter variational trial function, $m(\bu)= M \sqrt{\psi(\bu)}$, approaches our numerical prediction asymptotically in the limit of large filler fractions, while keeping a constant offset in the slender-rod limit. \cite{Finner2019PRL}
		We suppose that the asymptotic agreement of our analytical and numerical results is caused by a cancellation of errors between the Gaussian approximation and the solution of our variational theory, which, due to the choice of trial function, arguably neglects certain correlations that are implicit in the connectedness Ornstein-Zernike equation.

\begin{figure}
			\includegraphics[width = 0.95 \linewidth]{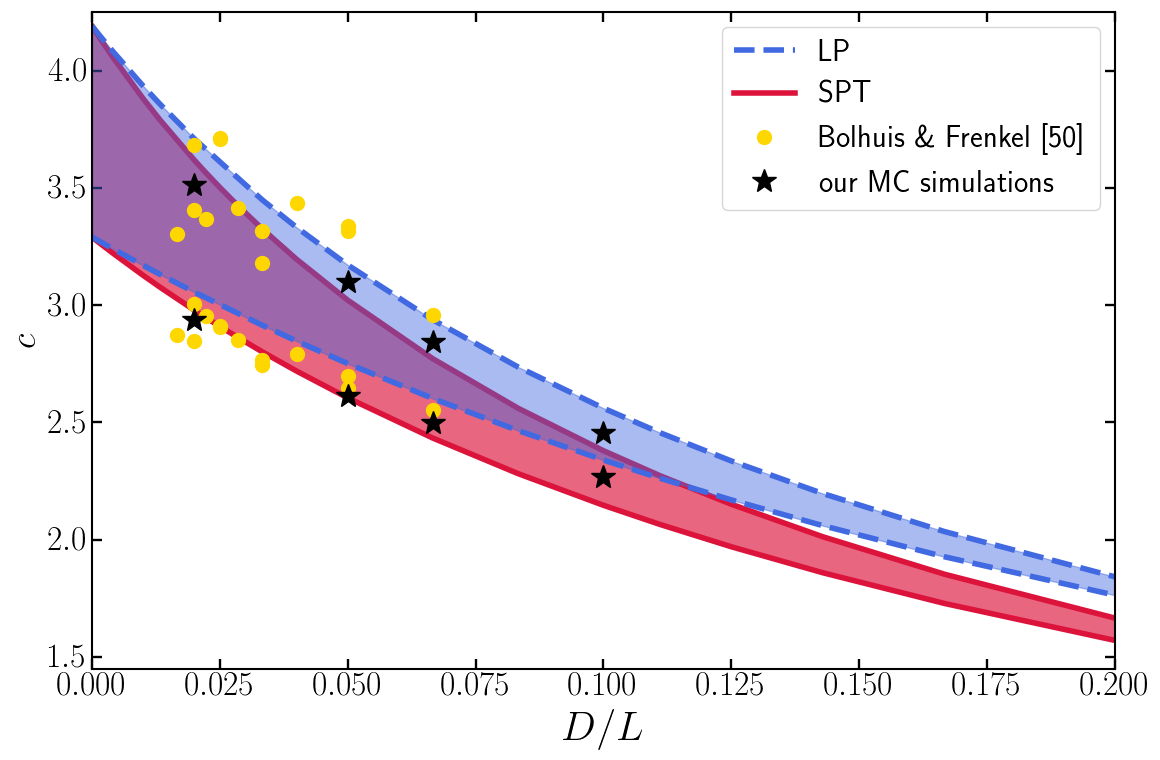}
			\caption{
				Shaded areas: isotropic-nematic coexistence regions as obtained by Lee-Parsons theory (LP, dashed-blue) \cite{Lee1987}, and Scaled Particle Theory (SPT, solid red) \cite{TuinierBook}.
				Monte Carlo simulation data are estimates from our own simulations (black stars), and the re-plotted results of Bolhuis and Frenkel \cite{Bolhuis1997} (yellow circles).
			}
			\label{fig:coexistence}
		\end{figure}

\begin{figure}
			\includegraphics[width = 0.95 \linewidth]{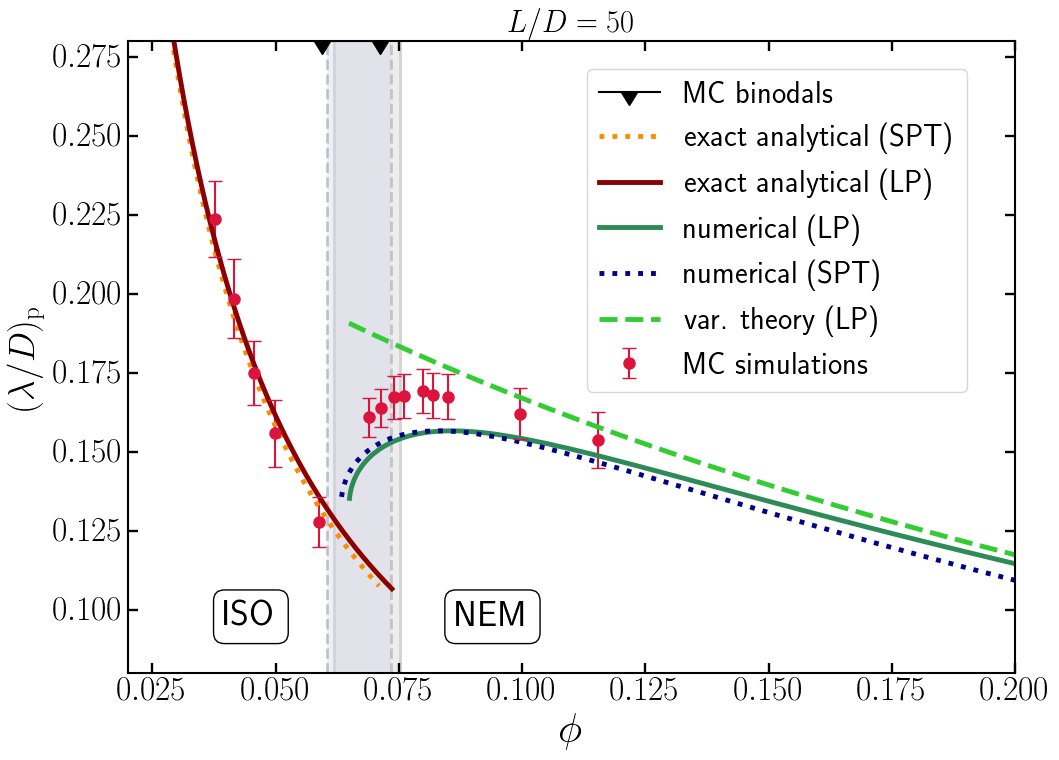}
			\caption{
				Critical shell thickness $\lambda_\pp/D$ for percolation \textit{vs.} the volume fraction $\phi$ for hard spherocylinders with aspect ratio $L/D=50$.
				The green and brown solid lines indicate our (numerically) exact results obtained with the Lee-Parsons approach (LP).
				The dotted lines those obtained using the approach based on Scaled Particle Theory (SPT).
				The results from our variational theory with the Lee-Parsons closure are shown by the light green dashed line.
				Our Monte Carlo simulation results are represented by red circles.
				Also indicated are the isotropic and nematic binodals (grey vertical lines) obtained from Lee-Parsons (solid) and the Scaled Particle Theory (dashed), see Figure \ref{fig:coexistence}. \cite{Lee1987, TuinierBook}
				The binodal estimates from our MC simulations are denoted by black carets at the top axis of the Figure.
				Note that the simulated dispersions with volume fractions close to or within the estimated coexistence regions do appear stable in our simulations, but might actually be metastable, see Section \ref{sec:numericsMC}.
			}
			\label{fig:PD50}
		\end{figure}

		To estimate the accuracy of our theory, we also compare in Figure \ref{fig:PD100} our theoretical predictions with the results of our MC simulations, as already published in \cite{Finner2019PRL}.
		We find excellent quantitative agreement between theory and simulations in the isotropic phase, as expected. \cite{Schilling2015}
		In the nematic phase, the agrement is still semi-quantitative, but the theory systematically underestimates the MC result by $5 - 8\%$.
		We surmise that this small discrepancy arises partly due to the finite size of the simulation box, leading to slightly higher nematic order parameters in the simulations, but also because an effective rescaling of the two-particle excluded and contact volume might be less accurate in the nematic than in the isotropic phase.
		Also indicated in the Figure are the volume fractions at phase coexistence according to Lee-Parsons theory, $\phi_\mathrm{iso} = 0.032$ and $\phi_\mathrm{nem} = 0.040$, between which our numerical solutions are metastable, as denoted by the grey shaded area (see also Figure \ref{fig:coexistence}). \cite{Lee1987}
		While our MC simulations in this region seem stable, possibly due to the finite size of the simulation box or the limited accuracy of Lee-Parsons theory, we cannot rule out the possibility of the simulated dispersion actually being metastable, see also the discussion in Section \ref{sec:numericsMC}.
		
		For clarity, we do not show our predictions following from the Scaled Particle Theory approach, Eq. \eqref{GammaSPT}.
		Our results for both approaches lie within $1\%$ of each other for the $\phi$-range of our MC simulations, and within $3\%$ in the entire plot range of Figure \ref{fig:PD100}, as the inset to Figure \ref{fig:PD100} shows.
		Note that the discrepancy between the two approaches does grow, and becomes significant at high volume fractions, as demonstrated in the inset to Figure \ref{fig:PD100}.
		In the limit $L \gg D$ and $\lambda \ll D/4$, \textit{i.e.}, in the high-density nematic phase of slender (but not infinitely slender) rods, our analytical theory predicts $\lambda_\pp / D \sim (\phi -1)^2 / x$, with $x=16$ for the Lee-Parsons closure and $x=128/3\approx 43$ for our new closure inspired by Scaled Particle Theory.
		The rods are in that case aligned almost perfectly, and the predictions from Lee-Parsons and Scaled Particle Theory deviate substantially. 
		Note that at such high volume fractions percolation in the nematic may be pre-empted by a transition to the smectic A phase. \cite{Bolhuis1997}

    		%% coexistence

		%%% 50 %%%
		
		%%% 20 %%%
		\begin{figure}
			\includegraphics[width = 0.95 \linewidth]{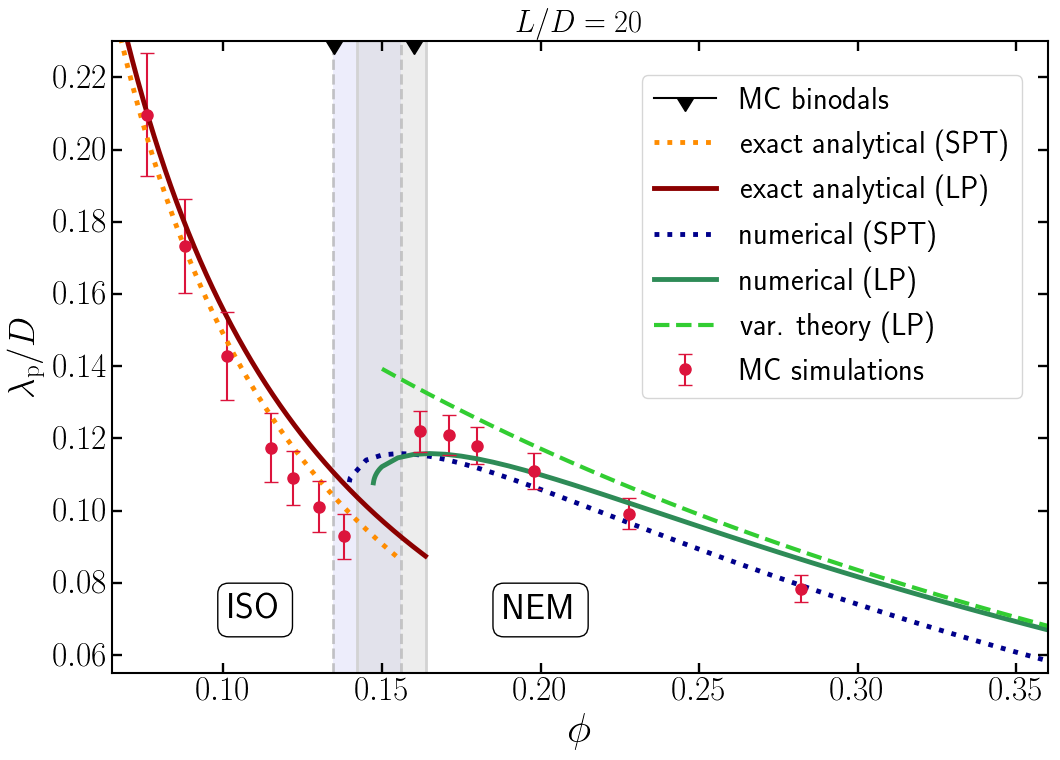}
			\caption{
				Critical shell thickness $\lambda_\pp/D$ for percolation \textit{vs.} the volume fraction $\phi$ for hard spherocylinders with aspect ratio $L/D=20$.
				The green and brown solid lines indicate our (numerically) exact results obtained with the Lee-Parsons approach (LP).
				The dotted lines are those resulting from the approach based on Scaled Particle Theory (SPT).
				The results from our variational theory with the Lee-Parsons closure are shown by the light green dashed line.
				Our Monte Carlo simulation results are represented by red circles.
				Also indicated are the isotropic and nematic binodals (grey vertical lines) obtained from Lee-Parsons (solid) and the Scaled Particle Theory (dashed), see also Figure \ref{fig:coexistence}. \cite{Lee1987, TuinierBook}
				The binodal estimates from our MC simulations are denoted by black carets at the top axis of the Figure.
				Note that the simulated dispersions with volume fractions close to or within the estimated coexistence regions do appear stable in our simulations, but might actually be metastable, see Section \ref{sec:numericsMC}.
			}
			\label{fig:PD20}
		\end{figure}

		%%% 10 %%%
		\begin{figure}
			\includegraphics[width = 0.95 \linewidth]{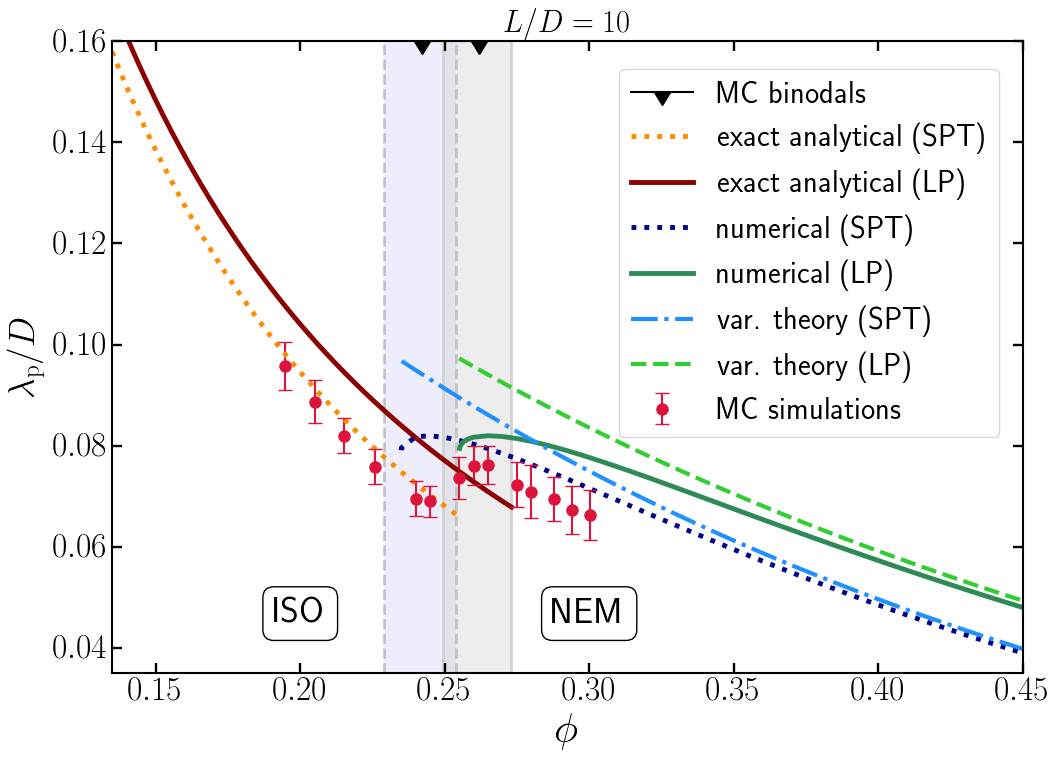}
			\caption{
				Critical shell thickness $\lambda_\pp/D$ for percolation as a function of the volume fraction $\phi$ for the aspect ratio $L/D=10$.
				The solid lines indicate our (numerically) exact results with the Lee-Parsons approach (LP),
				and the dotted lines those obtained using the approach based on Scaled Particle Theory (SPT).
				Predictions of our variational theory are indicated by the light green dashed line (LP) and by the blue dash-dotted line (SPT).
				Our Monte Carlo simulation results are represented by red circles.
				The isotropic and nematic binodals, represented by grey vertical lines, follow from Lee-Parsons theory (solid) and Scaled Particle Theory (dashed), respectively, see also Figure \ref{fig:coexistence}. \cite{Lee1987, TuinierBook}
				The binodal estimates from our MC simulations are denoted by black carets at the top axis of the Figure.
				Note that the simulated dispersions with volume fractions close to or within the estimated coexistence regions do appear stable in our simulations, but might actually be metastable, see Section \ref{sec:numericsMC}.
				We plot our theoretical predictions up to a volume fraction of $\phi \approx 0.45$, around which we expect a transition to the smectic-A phase. \cite{Bolhuis1997}
			}
			\label{fig:PD10}
		\end{figure}

		%%% 5 %%%
		\begin{figure}
			\includegraphics[width = 0.95 \linewidth]{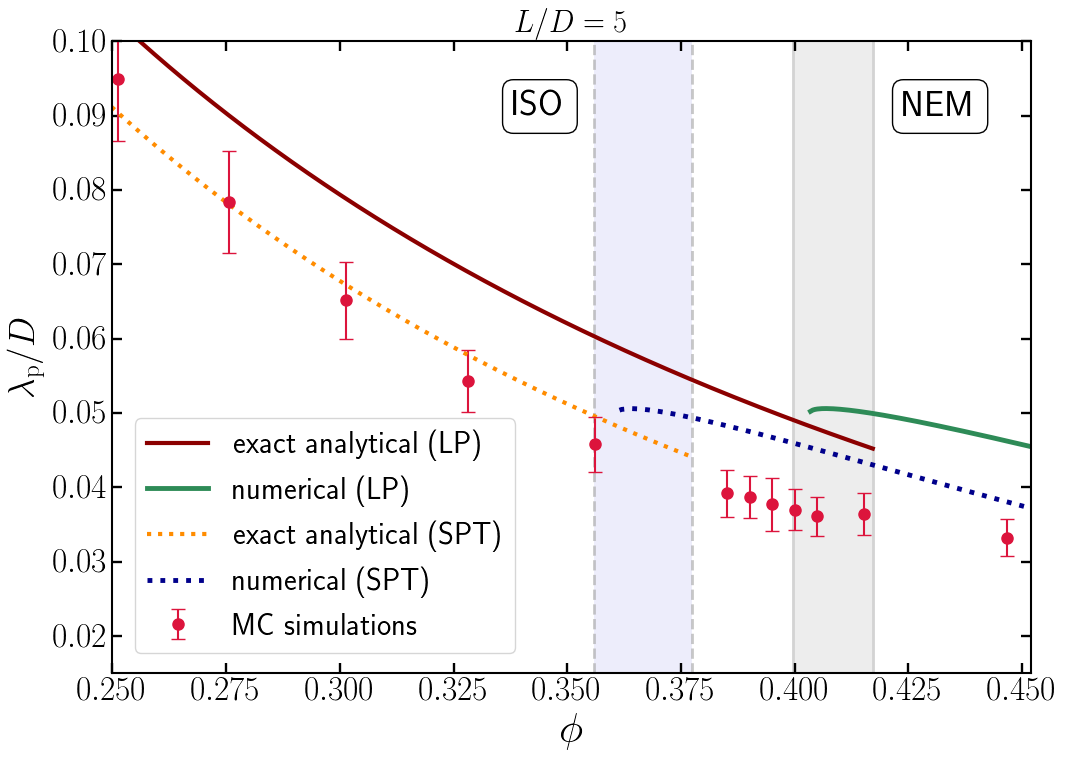}
			\caption{
				Critical shell thickness $\lambda_\pp/D$ at percolation as a function of the volume fraction $\phi$ for the aspect ratio $L/D=5$.
				The solid lines indicate our (numerically) exact results with the Lee-Parsons approach (LP),
				and the dotted lines those obtained using the approach based on Scaled Particle Theory (SPT).
				Our Monte Carlo simulation results are represented by red circles.
				The isotropic and nematic binodals, represented by grey vertical lines, follow from Lee-Parsons theory (solid) and Scaled Particle Theory (dashed), respectively, see also Figure \ref{fig:coexistence}. \cite{Lee1987, TuinierBook}
				Note that the simulated dispersions with volume fractions close to or within the theoretical coexistence regions do appear stable in our simulations, but might actually be metastable, see Section \ref{sec:numericsMC}.
				We plot our theoretical predictions up to a volume fraction of $\phi \approx 0.45$, around which we expect a transition to the smectic-A phase. \cite{Bolhuis1997}
			}
			\label{fig:PD5}
		\end{figure}

		As the difference between the predictions of Lee-Parsons and Scaled Particle Theory is much smaller than that between our theory and simulation results, we conclude that both scaling factors, Eq. \eqref{GammaLP} and \eqref{GammaSPT}, work about equally well for predicting the percolation behavior of hard spherocylinders of aspect ratio 100.
		For smaller aspect ratios, however, this state of affairs changes.
		To demonstrate this, we also calculate the orientational distributions and the percolation thresholds for rods of aspect ratio $L/D = 50$, $L/D = 20$, $L/D = 10$ and $L/D = 5$, respectively.
		Our calculations show that the percolation threshold for shorter rods exhibits similar trends as those in Figure \ref{fig:PD100}.
		However, with decreasing aspect ratio, we start to see significant \textit{quantitative} differences between the Lee-Parsons and the Scaled Particle Theory approach.
		
		One difference manifests itself in the location of the coexistence regions.
		For comparison, we plot in Figure \ref{fig:coexistence} the isotropic and nematic coexistence densities according to Lee-Parsons theory (LP) \cite{Lee1987} and Scaled Particle Theory (SPT) \cite{TuinierBook}.
    		Also indicated are the Monte Carlo simulation results by Bolhuis and Frenkel \cite{Bolhuis1997}, and the binodal estimates from our own MC simulations.
    		%The Figure shows that, for small aspect ratios $L/D$, the Lee-Parsons approach predicts the phase transition more accurately than Scaled Particle Theory does. \todo{does this still hold with the new data?}

		Focusing on the percolation threshold itself, as shown in the Figures \ref{fig:PD50}, \ref{fig:PD20}, \ref{fig:PD10} and \ref{fig:PD5}, we find that our new closure inspired by Scaled Particle Theory compares much better with our MC results than the Lee-Parsons approach, at least for short rods in the high-density isotropic and in the high-density nematic phase.
		In the isotropic phase of rods of aspect ratio $L/D=10$, for instance, using our new Scaled Particle closure instead of the Lee-Parsons closure results in a reduction of the largest relative discrepancy from our MC data from $\sim17\%$ to less than $5\%$, see Figure \ref{fig:PD10}.
		In the nematic phase, our Scaled Particle closure predicts a percolation threshold within $\sim8\%$ of the MC result, while the Lee-Parsons approach produces a deviation of $\sim17\%$.
		As mentioned earlier, the difference between the two closures is particularly pronounced for short rods, but becomes negligible for particles of aspect ratio $L/D\geq 100$.

%%%%%%%%%%%%%%%%%%%%%%%%%%%

	\section{Discussion and conclusions \label{sec:dis}}

		We have investigated the continuum percolation of hard spherocylinders in the uniaxial nematic phase of slender nanoparticles by means of connectedness percolation theory and Monte Carlo simulations, and focused attention on the impact of finite particle aspect ratio.
		In our earlier work we showed that percolation in the nematic phase of  infinitely slender rods cannot be achieved by increasing the particle density, and that adding more particles to the suspension may even lead to a decay of the percolating cluster -- a counterintuitive behavior, and opposite to what happens in the isotropic phase. \cite{Finner2019PRL}
		Here, we find that for nanofillers with aspect ratios below $\simeq20$ adding particles to a nematic suspension always \textit{aids} cluster formation:
		if the rods are sufficiently short, a density increase can no longer cause a loss of percolation-- a behavior similar to that in the isotropic phase, and contrary to our predictions in the slender rod limit. \cite{Finner2019PRL}
		For moderate aspect ratios between the two extremes of short and infinitely slender rods, we find a crossover of the two effects that results in re-entrance: percolation that takes place in the low-density nematic may be lost upon increasing the density, and obtained again by adding more particles to the suspension.

		To account for finite particle aspect ratios in our theoretical model and go beyond the second virial approximation, we propose a new closure relation inspired by Scaled Particle Theory.
		Similarly to the known Lee-Parsons closure, \cite{Schilling2015} our new Scaled Particle closure approximately accounts for higher order virials by effectively rescaling the contact volume of particle pairs.
		Comparison with the results of our MC simulations demonstrates that our new closure provides a more accurate prediction for the percolation threshold than the Lee-Parsons approach does, even though the latter seems to predict the I-N transition of short spherocylinders more accurately. \cite{Lee1987, TuinierBook}
		
		The advantage of our approach based on Scaled Particle Theory compared to the Lee-Parsons closure turns out to be particularly pronounced at high densities in both the isotropic and nematic phase, and for modest particle anisometries.
		We surmise that it arises because the weights of the higher-order interaction terms in the (diagrammatic) expansion of the direct \textit{correlation} function are different to those in the expansion of the direct \textit{connectedness} function.\,\cite{Coniglio1977, Sevick1990}
		As a result, the diagrams governing the phase behavior of spherocylinders are slightly different to those determining their percolation threshold.
		Both the Lee-Parsons and the Scaled Particle Theory routes produce an effective scaling factor $\Gamma(\phi, L/D)$ that accounts for these higher order diagrams in an approximative way, but with different weights for each of the terms.
		
		In addition to solving the underlying equations for both closures numerically by recursive iteration, we applied a Schwinger-type variational percolation theory and combined it with the Gaussian approximation of the orientational distribution function. \cite{Odijk1986}
		An advantage of this type of theory is that its predictions may be systematically improved by using trial functions $m(\bu)$ with an increasing number of variational parameters.
		The solution obtained for our choice of trial function with one variational parameter, $m(\bu)=M\sqrt{\psi(\bu)}$, turns out to provide very accurate predictions for the percolation threshold deep in the nematic phase, and approaches our numerical solution asymptotically, at least for finite aspect ratios. \cite{Finner2019PRL}
		
		Using a two-parameter trial function of the form $m(\bu)=\sqrt{\psi(\bu)}\big(M+ N \theta^2\big)$, with $\theta$ the angle between the nematic director and the particle orientation vector $\bu$, generally increases the accuracy of our analytical prediction. 
		However, the improvement is significant only for nanoparticles of very large aspect ratio in the low-density nematic.
		For the aspect ratios studied in this article, the relative difference (at most 4\% for $L/D=100$ and 2.1\% for $L/D=20$) is rather minor, considering that our analytical prediction entirely misses the nonmonotonic re-entrance behavior of the percolation threshold.
		The reason for the latter is the Gaussian approximation, which loses accuracy close to the nematic melting transition. \cite{Odijk1986}. 
		In fact, if we use the \textit{exact} distribution function for $\psi(\theta)$ instead of the Gaussian approximation, our variational theory proves to be very powerful, predicting a percolation threshold within $0.82\%$ of the numerically exact solution in the entire density range of the nematic (for the two-parameter trial function and $L/D\rightarrow\infty$).

		The question arises how cluster formation in suspensions of anisometric particles changes for different particle shapes, and in other symmetry-broken phases.
		For hard \textit{platelets}, the percolation threshold in the uniaxial nematic has recently been investigated by MC simulations and turns out to decrease with the particle concentration,  \cite{TanjaPlatelets2012} similar to our findings in this article for hard rods of moderate aspect ratio.
		However, it is not yet clear if percolation in a platelet suspension can also be \textit{lost} with increasing density in a similar fashion as in a rod suspension.
		It also remains to be seen how percolation of elongated particles occurs in other liquid crystal phases, where both the orientational and the translational symmetries are broken.
		In the smectic phases, for example, it might be possible to obtain a percolating network in two directions in the plane of the smectic layer, while the normal direction remains disconnected (depending on the layer spacing).
		This we intend to address in a forthcoming publication.

%%%%%%%%%%%%%%%%%%%%%%%%%%%%%%%%%%%%%%%%%%%%%%%%%%%%%%
	\begin{acknowledgments}
		S.\,P.\,F.\ and P.\,v.\,d.\,S.\ are funded by the European Union's Horizon 2020 research and innovation programme under the Marie Sk\l{}odowska-Curie grant agreement No 641839.
		Computer simulations presented in this paper were carried out using the bwForCluster NEMO high-performance computing facility.
	\end{acknowledgments}

	\appendix
    	\section{The direct correlation function within Scaled Particle Theory} \label{app:A}
    	
    		The excess free energy (over the ideal gas) of a uniform particle suspension has the form \cite{Hansen}
    		\begin{align}
    			\frac{\beta F_\text{exc}}{N} &= \frac{n}{2}  \int \dd \mathbf{r} \langle \langle C (\mathbf{r}, \bu, \bu') \rangle \rangle' 
    				=  \frac{n}{2} \langle \langle \hat{C} (\bu, \bu') \rangle \rangle' . \label{fexc}
    		\end{align}
    		Using a generalized closure relation,
    		\begin{align}
    			C(\br, \bu, \bu') = \Gamma(n, L, D) f(\br, \bu, \bu') + \Omega(\br, n, L, D),
    		\end{align}
    		and Eq. \eqref{vex}, this excess free energy can be re-written as
    		\begin{align}
    			\frac{\beta F_\text{exc}}{N} =c\,& \Gamma(n, L, D) \rho[\psi(\bu)] + \Gamma(n, L, D) \frac{n}{2} G \nonumber \\
    			 	&- \frac{n}{2} \hat{\Omega}(n, L, D) . \label{fexc}
    		\end{align}
    		Here, $c=n \pi L^2 D / 4 = \phi (D/L+2D^2/3L^2)^{-1}$ is the dimensionless particle concentration, $\rho[\psi(\bu)] \equiv 4 \langle \langle |\bu \times \bu' \rangle \rangle' / \pi$, and $G \equiv 2\pi L D^2 + 4 \pi D^3 / 3$ denotes the contribution of the hemispherical end-caps
    		to the two-particle excluded volume.
\onecolumngrid
\vspace{\columnsep}
    		The operator $\hat{(\cdots)} = \int\dd\br (\cdots)$ again denotes a volume integral and is equivalent to a zero wave vector Fourier transform.\\
    		
    		The goal is now to determine the functions $\Gamma(n, L, D)$ and $\hat{\Omega}(n, L, D)$ of our generalized closure relation by using the frameworks of Lee-Parsons theory and Scaled Particle Theory.
    		Within Lee-Parsons theory, the excess free energy is given by \cite{Lee1987}
    		\begin{align}
    			\frac{\beta F_\text{exc, LP}}{N} &= \frac{1}{8} \frac{\phi(4-3\phi)}{(1-\phi)^2} \left(8 + 3\left( \frac{L}{D}\right)^2 \frac{\rho[\psi(\bu)]}{1 + 3L/2D}\right), \label{fexcLP}
    		\end{align}
    		
    		with $\phi = n \pi D^2 [3L + 2D]/12$ the volume fraction of particles in the suspension.
    		Comparison with Eq. \eqref{fexc} yields the familiar result
    		\begin{align}
    			\Gamma(n, L, D) = \Gamma_\text{LP}(\phi) = \frac{1 - 3\phi / 4}{(1-\phi)^2} \hspace{1cm} \text{and} \hspace{1cm} \hat{\Omega}(n, L, D) = 0.
    		\end{align}
    		Within Scaled Particle Theory, the excess free energy takes the form \cite{TuinierBook}
    		\begin{align}
    			\frac{\beta F_\text{exc, SPT}}{N} &= -\ln(1-\phi) + A[\psi(\bu)]\frac{\phi}{1-\phi} + \frac{B[\psi(\bu)]}{2}\frac{\phi^2}{(1-\phi)^2}.
    		\end{align}
    		Here,
    		\begin{gather}
    			A[\psi(\bu)] = 3 + \frac{3}{2 + 3L / D}\left(\frac{L}{D}\right)^2 \, \rho[\psi(\bu)]\\
    			\shortintertext{and}
			B[\psi(\bu)]= 12\frac{(1 + L / D)(1 + 2 L / D)}{(2 + 3L / D)^2} + 12\left( \frac{L}{D}\right)^2\frac{1 + L / D}{(2 + 3L / D)^2} \rho[\psi(\bu)].
    		\end{gather}
    		Separating the orientation-dependent and -independent terms yields the excess free energy
    		\begin{align}
    			\frac{\beta F_\text{exc, SPT}}{N} &= X + \rho[\psi(\bu)] \Big[ \frac{\phi}{1-\phi} \left( \frac{L}{D}\right)^2  \frac{3}{2 + 3L / D}+ \frac{6\phi^2}{(1-\phi)^2} \left( \frac{L}{D}\right)^2 \frac{1 + L / D}{(2 + 3L / D)^2} \Big],
    		\end{align}
    		with
    		\begin{align}
    			X = -\ln(1-\phi) + \frac{3\phi}{1-\phi} + \frac{6\phi^2}{(1-\phi)^2} \frac{(1 + L / D)(1 + 2 L / D)}{(2 + 3L / D)^2}.
    		\end{align}
    		Comparison with Eq. \eqref{fexc} shows that, within the framework of Scaled Particle Theory,
    		\begin{align}
    			\Gamma(n, L, D) = \Gamma_\text{SPT}(\phi, L/D) =  (1-\phi)^{-1} \Big[ 1 +  \frac{\phi}{1-\phi}\frac{2+ 2 L / D}{2 + 3L / D} \Big],
    		\end{align}
    		and
    		\begin{align}
    			\hat{\Omega}(n, L, D)  &= G \Gamma(\phi, L/D) - \frac{2X}{n}\\
    				&=\frac{2}{n} \Big[\frac{\phi}{1-\phi} + \ln(1-\phi) + \frac{\phi^2}{(1-\phi)^2}\frac{2 (1 + L / D)}{2 + 3L / D} \left(4 - 3\frac{1 + 2 L / D}{2 + 3L / D}\right) \Big].
    		\end{align}
    		%\end{widetext}
\vspace{\columnsep}
\twocolumngrid

	\bibliography{lit}

\end{document}